\documentclass[12pt,preprint]{aastex}

\shorttitle{Numerical simulations of HH~555}
\shortauthors{Kajdi\v{c} \& Raga}

\begin{document}


\title{Numerical simulations of HH~555}


\author{P. Kajdi\v{c}}
\affil{Instituto de Astronom\' ia, Universidad Nacional Aut\'onoma de M\'exico,
Apartado Postal 70 - 264, Ciudad Universitaria,  M\'exico, D.F., CP 04510}
\email{primoz@astroscu.unam.mx}

\and

\author{A. C. Raga}
\affil{Instituto de Ciencias Nucleares, Universidad Nacional Aut\'onoma de M\'exico,
Apartado Postal 70-543, Ciudad Universitaria,  M\'exico, D.F., CP 04510}


\begin{abstract}
We present 3D gasdynamic simulations of the Herbig Haro object
HH~555. HH~555 is a bipolar jet emerging
 from the tip of an elephant trunk entering the Pelican Nebula
from the adjacent molecular cloud. Both beams of HH~555 are curved 
away from the center of the H~II region. This indicates that
they are being deflected by a side-wind probably coming from
a star located inside  the nebula or by the expansion of the
nebula itself. HH~555 is most likely an irradiated jet emerging
from a highly embedded protostar, which has not 
yet been detected.

In our simulations we vary the incident photon flux, which
in one of our models is equal to the flux coming from a star 1 pc away
emitting 5$\times$10$^{48}$ ionizing (i.~e., with energies above the
H Lyman limit) photons per second. An external, plane-parallel flow
(a ``side-wind'') is coming from the same direction as the photoionizing flux.
We have made four simulations, decreasing the
photon flux by a factor of 10 in each simulation. We discuss the
properties of the flow and we compute H$\alpha$ emission maps
(integrated along lines of sight).
 We show that the level of the incident photon flux has an important 
influence on the shape and visibility
of the jet. If the flux is very high, it causes a strong evaporation 
of the neutral clump, producing a photoevaporated wind traveling in the 
direction opposite to the incident flow. The interaction of the two flows
creates a double shock ``working surface'' around the clump protecting it and
the jet from the external flow. The jet only starts to curve when it penetrates
through the working surface.
\end{abstract}





\section{Introduction}
Many new Herbig Haro objects have been discovered in recent years 
by various deep imaging surveys (i.~e. Bally \& Reipurth 2001, 2003, Reipurth et al. 2004 
and Walawander, Bally \& Reipurth 2005). These objects come in different forms and shapes. 
Some of them are seen as barely observable spots, others appear in the form of chains of 
knots or as symmetric or asymmetric jets. The jets can be straight or curved
(see for example the jets described
in Bally \& Reipurth 2001). The curving of the jets is usually interpreted as
the interaction of the jet with an enviorment in relative motion with respect to the outflow
source. Numerical simulations have been made for the curved
jets in the Orion Nebula (M42) by Masciadri \& Raga (2001). The problem of a jet
in a sidewind  has also been aproached analytically and experimentally. Cant\'o \& Raga (1995) found 
an analytical solution for the shape of the jet, while Lebedev et al. (2004)
studied jet deflection by crosswinds in laboratory.

In the present paper, we simulate a peculiar Herbig Haro object known as
HH~555 located in the Pelican Nebula (IC4050), discovered by Bally \& Reipurth 
(2003) (see Figure~1). HH~555 is a bipolar jet emerging
from the tip of an elephant trunk embedded in the nebula. The jet and
counterjet both curve away from the ionizing source, indicating that
they might be interacting with a stellar wind or with the expanding
H~II region.

We carry out 3D gasdynamical simulations in which we include the ionizing
photon flux (from $\theta$ Orionis) and the expanding H~II region (modeled
as a plane-parallel flow entering the computational grid in a direction
parallel to the impinging ionizing photon flux) interacting
with the elephant trunk (modeled as a dense cylinder+spherical cap,
neutral structure). From close to the tip of the elephant trunk, a bipolar
outflow is injected, with an axis at an angle to the impinging, ionizing
flux. We present the results for four models exploring different values
for the impinging ionizing photon flux (but with otherwise identical
parameters), and compare the results in a qualitative way with the
observations of HH~555.

The paper is organized as follows. In section 2, we describe the properties
of HH~555, as derived by Bally \& Reipurth (2003). In section 3, we present
the numerical simulations, and predictions of H$\alpha$ emission maps
are discussed in section 4. Finally, the conclusions are given in section 5.

\section{Physical properties of HH~555}
Bally \& Reipurth (2003) give the following description of the elephant trunk and HH~555:
the elephant trunk is a $4'$ long filament of neutral gas, which penetrates into the Pelican Nebula. It 
shows a dense condensation at its tip, with
a $\sim 20''$ diameter. Two curved flows emerge from it in a direction almost perperdicular 
to the elephant trunk. They are the southern and the northern outflow lobes,
which have lengths of $\sim 25''$. The flows bend at an angle 
of 25$^\circ$ in the direction
away from the center of the Pelican Nebula, giving it a ``C''-shaped
symmetry. The angular width of the jets is of $\approx 2''$, which for
an assumed distance of 600 pc implies jet diameters of 1200 AU.

The radial velocity measurements of the southern jet (the spectra
for the northern jet have not 
been obtained yet) give values of up to $-80$~km~s$^{-1}$ with respect to
the background emission of the Pelican Nebula. From the [S~II] doublet ratio,
$I(\lambda 6717)/I(\lambda 6731)$, the electron density was estimated
to be $n_e \approx$600 cm$^{-3}$ in
contrast to $n_e \approx$200 cm$^{-3}$ for the background nebula. 
From these data the mass-loss rate is estimated to be
$\dot{M} = 1.4 \times 10^{-7}$ M$_{\odot}$~yr$^{-1}$. 
The source of the jets is probably highly embedded inside the
elephant trunk, since it could not be detected at infrared wavelengths.

\section{The Numerical Simulations}

In order to simulate the HH~555 outflow, we have carried out 3D numerical
simulations in which we include a number of elements, which are shown
in the schematic diagram of Figure~2. These elements are:
\begin{itemize}
\item a plane-parallel ionizing photon flux $F_*$ which enters the
computational grid in the $+x$-direction. We compute 4 models (models~1
through 4) with $F_*=(4.2,42,420,4200)\times 10^{7}$~cm$^{-2}$s$^{-1}$
(respectively). The highest of these fluxes corresponds to an
unshielded ionizing
source with an ionizing photon rate of $S_*=5\times 10^{48}$~s$^{-1}$ located
at a distance of 1~pc from the center of the computational grid. The
lower values of $F_*$ could be the result of larger distances to the ionizing
photon source, or could be interpreted as the effect of absorption (due to
dust or to hydrogen photoionization) in the material between the source
and the simulated region. In all models, a black body spectral
distribution with a $T_*=40000$~K temperature is assumed,
\item an expanding photoionized region, modeled as a plane-parallel inflow
(also in the $+x$-direction) of gas with a $v_w=50$~km~s$^{-1}$ velocity,
$n_w=10$~cm$^{-3}$ density and $T_w=10^4$~K temperature. The expanding
H~II region is identical in the four computed models,
\item an ``elephant trunk'', modeled as a neutral,
initially stationary cylinder+spherical
cap structure aligned with the $x$-axis, with a $r_c=7.56\times 10^{16}$~cm
radius. This neutral structure has a $n_c=10^4$~cm$^{-3}$ density
and $T_c=100$~K temperature,
\item a bipolar outflow with an initial radius $r_j=3\times 10^{15}$~cm,
velocity $v_j=100$~km~s$^{-1}$,
density $n_j=600$~cm$^{-3}$ and temperature $T_j=1000$~K. The outflow
is imposed in a cylinder of radius $r_j$ and length $2r_j$ (with
velocities of $\pm v_j$ along the symmetry axis of this cylinder, in
the top and bottom halfs of the cylinder, respectively).
The bipolar outflow is initially neutral. The center of
the cylinder coincides with the center of the spherical cap (which
forms the end of the neutral elephant trunk, see above), and also with the
center of the $(x,y,z)$ coordinate system. The outflow
axis lies on the $xz$-plane and forms an angle $\alpha=70^\circ$ with the
$x$-axis (see Figure~2).
\end{itemize}

The numerical simulations are computed with the ``yguaz\'u-a'' adaptive grid
code (see the description of Raga, Navarro-Gonz\'alez \& Villagr\'an-Muniz
2000), using the version of the code described by Raga \& Reipurth (2004).
This version of the code solves the 3D gasdynamic equations together with
a single rate equation for neutral hydrogen. Radiative recombination,
and collisional and photo-ionizations are included. A simultaneous solution is
made of the transfer of ionizing photons (at the Lyman limit) along the
$x$-axis. A parametrized cooling function (calculated as a function of
the ionization fraction, density and temperature) and the heating due
to photoionization of H are also included in the energy equation (see Raga
\& Reipurth 2004). We should note that because of our ``Lyman limit
only'' approximation to the radiative transfer, the spectral distribution
(which we assume to be a black body with $T_*=40000$~K, see above)
only appears in the photoionization heating term (see Cant\'o et al.
1998).

From the results of our simulations, we compute H$\alpha$ emission maps
by integrating the H$\alpha$ emission coefficient along lines of sight.
We compute this emission coefficient by adding the contributions
of the ``case B'' recombination cascade and the collisional excitations
from the $n=1$ level of H.

For our computations, we have chosen a 5-level, binary adaptive grid
with a maximum resolution of $1.27\times 10^{15}$~cm (along the
three axes). The computational domain has an extent of
$(6.5,3.25,6.5)\times 10^{17}$~cm
along the $(x,y,z)$-axes (respectively). Transmission
conditions are imposed on all boundaries except for the $-x$ boundary,
in which the ``expanding H~II region'' inflow condition (see above)
is imposed.

In our simulations, the expanding H~II region (with a $n_w=10$~cm$^{-3}$
density and a $v_w=50$~km~s$^{-1}$ velocity) forms a bow shock
against the photoevaporated wind from the elephant trunk. From
the models of Hartigan et al. (1987), we find that a plane-parallel,
stationary shock model with a preshock density of 10~cm$^{-3}$
and a 50~km~s$^{-1}$ shock velocity has a cooling distance (to
$10^4$~K) $d_c\approx 6\times 10^{14}$~cm. At the head of the leading
bow shock of the jet, we have a shock velocity $\sim v_j=100$~km~s$^{-1}$,
and a preshock density $>10$~cm$^{-3}$ (corresponding to the shocked
expanding H~II region). For a shock velocity of 100~km~s$^{-1}$
and a 10~cm$^{-3}$ preshock density, the models of Hartigan et al. (1987)
give a cooling distance (to $10^4$~K) of $1.2\times 10^{15}$~cm.
These estimates show that the post-shock cooling distances are not
resolved in our simulations. However, the contribution of these
regions to the H$\alpha$ emission of the photoionized flow should be
relatively small (see, e.~g., Masciadri \& Raga 2004).

The simulations start with the expanding H~II region occupying
all of the domain, except for the region with the elephant trunk.
The embedded outflow is also ``turned on'' at this initial time. 
The simulations proceed to a total integration time 3000 years.

\section{Results}

The results of our simulations are shown in Figures~3 and 4. We have
calculated four models with basically identical initial setup, except for the
incident photon flux. Our goal was to determine the influence of the flux on
the shape and visibility of the jet. The values of the photon flux are
$F_*=(4.2,42,420,4200)\times 10^{7}$~cm$^{-2}$s$^{-1}$ for models~1 through 4
(respectively). The photon flux in Model~4 is equal to the flux coming from a
star 1 pc away emitting 5$\times$10$^{48}$ ionizing photons per second (which
corresponds to an ionizing source such as $\theta$ Orionis). The lower values
of $F_*$ could be the result of the interstellar absorption in the material
between the source and the simulated object or could be a consequence of
larger distances to the ionizing photon source.

\subsection{The neutral clump}

Figure~3 shows the density stratifications on
the $xz$-plane (i.~e., the plane on which lies the outflow axis,
see \S~3) after a $t=3000$~yr integration time for models~1 through
4. These stratifications show the shape of the
bipolar outflow and of the eroded neutral clump.

In models~1 and 2, the expanding H~II region (impinging on the
neutral clump along the $x$-axis) forms a bow shock against
the neutral clump, and against the side of the outflow which
emerges from the clump. The clump material which is photoionized
by the impinging ionizing photon flux is confined to a very narrow
region (surrounding the neutral clump) by the expanding H~II region.
The shape of the clump changes because of the interaction with
the expanding H~II region and the ionizing photon flux.
Its dimension along the $x$-axis is
reduced as the time-integration progresses.

As the impinging photoionizing flux is
increased (models~3 and 4), so does the amount of photoevaporated
material. In models~3 and 4 the interaction of the photoevaporated wind 
with the expanding H~II region forms a detached two-shock structure.
In model~3, and more dramatically in model~4, the neutral structure
is eroded as it ejects a photoevaporated wind. This photoevaporation
is stronger in model~4, and for this model we already see the
axial collapse (always obtained in simulations of the
photoevaporation of neutral clumps, see, e.~g., Mellema et al. 1998)
at a $t=3000$~yr integration time.

The results from our models can be interpreted in terms of
the analytic photoevaporated wind/expanding H~II region interaction
model of Henney et al. (1996) and Raga et al. (2005). These
authors define a ``high ionizing photon flux'' and a
``low ionizing photon flux'' regime. In the first regime a
detached, two-shock structure is created by the interaction
of the photoevaporated wind with the impinging H~II region,
while in the second regime the photoionized clump material
is confined to a thin shell around the neutral clump.

In the ``high ionizing photon flux'' regime, the
distance (along the symmetry axis) between the two-shock structure
and the center of the neutral clump is
\begin{equation}r_w=\lambda^{1/2}r_0,\end{equation}
where $r_0$ is a radius of the neutral structure. The dimensionless
parameter $\lambda$ is given by
\begin{equation}\lambda=\frac{F_0c_I}{n_wv^2_w},\end{equation}
where $F_0$ is the photon flux which actually arrives to
the surface of the neutral clump at a given time, $c_I$ is the
sound speed of the photoevaporated wind ($\sim$10~km~s$^{-1}$),
and $n_w$ and $v_w$ are the density and the velocity of the impinging
expanding H~II region.

For $\lambda>1$, the two-shock structure is detached and the
flow is in the ``high ionizing photon flux'' regime. 
For  $\lambda<1$, the flow is in the ``low ionizing photon flux'' regime.
In the case of our models, we calculated analytical values of
$\lambda$ using the model of Raga et al. (2005), obtaining
$\lambda=0.14$, 0.80, 3.35 and 8.32 for models~1 through
4 (respectively). Consistently with these $\lambda$ values,
in our numerical simulations we obtain detached shock structures
only for models~3 and 4. 

In Figure~4 we show H$\alpha$ emission maps (obtained
by integrating the H$\alpha$ emission coefficient along the $y$-axis,
without considering the extinction within the elephant trunk).
In the case of models~1 and 2, very little emission
is seen from the region of interaction between the clump and the
impinging ionizing photon flux and expanding H~II region.
For these two models, the emission is dominated by the jet (described
in \S 4). The converse is true for models~3 and 4, in which the H$\alpha$
emission is dominated by the photoevaporated wind
and the region of interaction between the
photoevaporated wind and the expanding H~II region.

\subsection{The jet}

The shape of the jet can be seen in the density
stratifications shown in Figure~3. 
In the case of models~1 and 2, once outside the neutral clump,
the outflow lobes have different shapes. In both lobes,
the jet begins to curve as soon as it exits the neutral elephant trunk.
The bottom lobe (travelling in the $-z$ direction) has a much stronger
curvature, changing its direction of propagation from the $-x$ to the
$+x$ direction.

In the case of the models~3 and 4, the jets emerge from the neutral region
with basically unchanged directions of propagation. This is due to the
fact that the downward directed lobe is travelling within a region
occupied by the photoevaporated
wind (ejected from the neutral structure), which has a direction of
motion which is approxmately parallel to the jet. The upward directed
jet is travelling into a low density region, which is shielded from
the expanding H~II region by the detached, two-shock interaction
region.

Further out from
the outflow source, the jets interact with the two-shock
structure (resulting from the photoevaporated wind/expanding
H~II region interaction). In this interaction, the downward
directed jet curves substantially. The upward directed jet
curves in a less substantial way as it penetrates the
two-shock interaction region.

The H$\alpha$ emission maps (Figure~4) show that
in models~1 and 2 the emission is dominated by the
region in which the expanding H~II region interacts with the
jet beam. The emission has a strong resemblance to the jet/sidewind
interaction models of Masciadri \& Raga (2001).

In models~3 and 4, the H$\alpha$ emission is dominated by the
photoevaporated wind and the wind/expanding H~II region interaction,
two-shock structure (see \S 3 and Figure~4). For model~4,
only the upwards directed jet is visible where it penetrates the
two-shock structure (at $z\sim 2\times 10^{16}$~cm). For model~3,
both outflow lobes are visible, superimposed on the wings of the
two-wind interaction bow shock.

\section{Conclusions}

In this paper, we simulate the peculiar Herbig Haro object 
HH~555, located in the Pelican Nebula (IC4050), which is thought to
be irradiated by nearby stars and deflected by an expanding H~II region.

We computed four 3D simulations of a bipolar jet emerging from the tip
of a neutral ``elephant trunk''. The elephant trunk is aligned with
an impinging ionizing photon flux and a wind of ionized material
(i.~e., the expanding H~II region). The bipolar outflow emerges at
an $\alpha=70^\circ$ angle with respect to the axis of the elephant
trunk. All of these elements are taken directly from the observations
of HH~555 of Bally \& Reipurth (2003).

In our models, we explore the effect of varying the impinging
ionizing photon flux $F_*$ (while keeping constant all of the
other model parameters). We have computed 2 models (models~1 and 2)
in the ``low ionizing flux'' regime (see Henney et al. 1996), in which
the expanding H~II region forms a bow shock against the surface
of the neutral structure, and two models in the ``high ionizing
photon flux'' regime (models~3 and 4),
in which a detached, two-shock structure
is formed as a result of the interaction between a photoionized
wind and the expanding H~II region.

In models~1 and 2, the expanding H~II region directly interacts
with the bipolar outflow, starting at the point in which the
jets emerge from the neutral elephant trunk. The curvature of the
jets in these two models is similar to the one observed in HH~555,
and the predicted H$\alpha$ emission maps (see Figure~4)
have a strong qualitative resemblance to the observed object.

In models~3 and 4, we obtain a strong, photoevaporated wind bow
shock which shields the jets (emerging from the elephant
trunk) from the impinging expanding H~II region. The jets only
curve when they reach the photoevaporated wind bow shock. The
H$\alpha$ intensity maps are dominated by the emission of this
bow shock, and the jets only become visible when they start
penetrating through the bow shock wings. The obtained morphologies
(see Figure~4) do not resemble the observed structure of HH~555 (see Figure~1).

From this, we conclude that the observed morphology of HH~555
implies that the interaction between the expanding
H~II region+impinging ionizing photon flux and the elephant trunk
must be in the ``low ionizing photon flux'' regime of Henney
et al. (1986) (i.~e., the parameters are such that $\lambda<1$,
see equation~2). Provided that this condition is met, the models
do produce structures that resemble the observations of HH~555.

It is interesting to note that the H$\alpha$ intensity
maps obtained from models~3 and 4 (our ``high ionizing photon flux''
regime models) have a quite striking resemblance
to the emitting structure around LL~Orionis (see Bally, O'Dell
\& McCaughrean 2000). As LL~Orionis is not embedded in an elephant
trunk, our models therefore do not directly apply to this object.
However, as pointed out by Bally et al. (2000), a photoevaporated
wind from a neutral envelope surrounding the star appears to
be present in LL~Orionis.

The present work should be considered as a first exploration of
the problem of a bipolar outflow which emerges from the interior
of an externally photoionized neutral structure. Given the number
of different elements which form part of the problem, the immense
resulting parameter space (with many weakly constrained parameters)
is likely to result in a variety of flow morphologies which might
be appropriate for modelling different objects.

In particular, we set out to model HH~555, and we do find models which
reproduce the observed images of this object in a qualitative
way. Also, we find that some of our models have a stronger
resemblance to the outflow system ejected by LL~Orionis. Future
comparisons of model predictions with spectroscopic and proper motion
observations of these objects will be necessary in order to
test the models (and also to constrain the model parameters).

\acknowledgments
This work was supported by the CONACyT
grants 43103-F and 46828-F, the DGAPA (UNAM) grant IN~113605
and the ``Macroproyecto
de Tecnolog\'\i as para la Universidad de la Informaci\'on y la
Computaci\'on'' (Secretar\'\i a de Desarrollo Institucional de la UNAM,
Programa Transdisciplinario en Investigaci\'on y Desarrollo
para Facultades y Escuelas, Unidad de Apoyo a la Investigaci\'on en
Facultades y Escuelas). Primo\v{z} Kajdi\v{c} acknowledges the Direcci\'on
General de Estudios de Posgrado
of the UNAM for a scholarship supporting his graduate studies.
We thank an anonymous referee for helpful comments.

\clearpage

\begin{figure}
\epsscale{.80}
\plotone{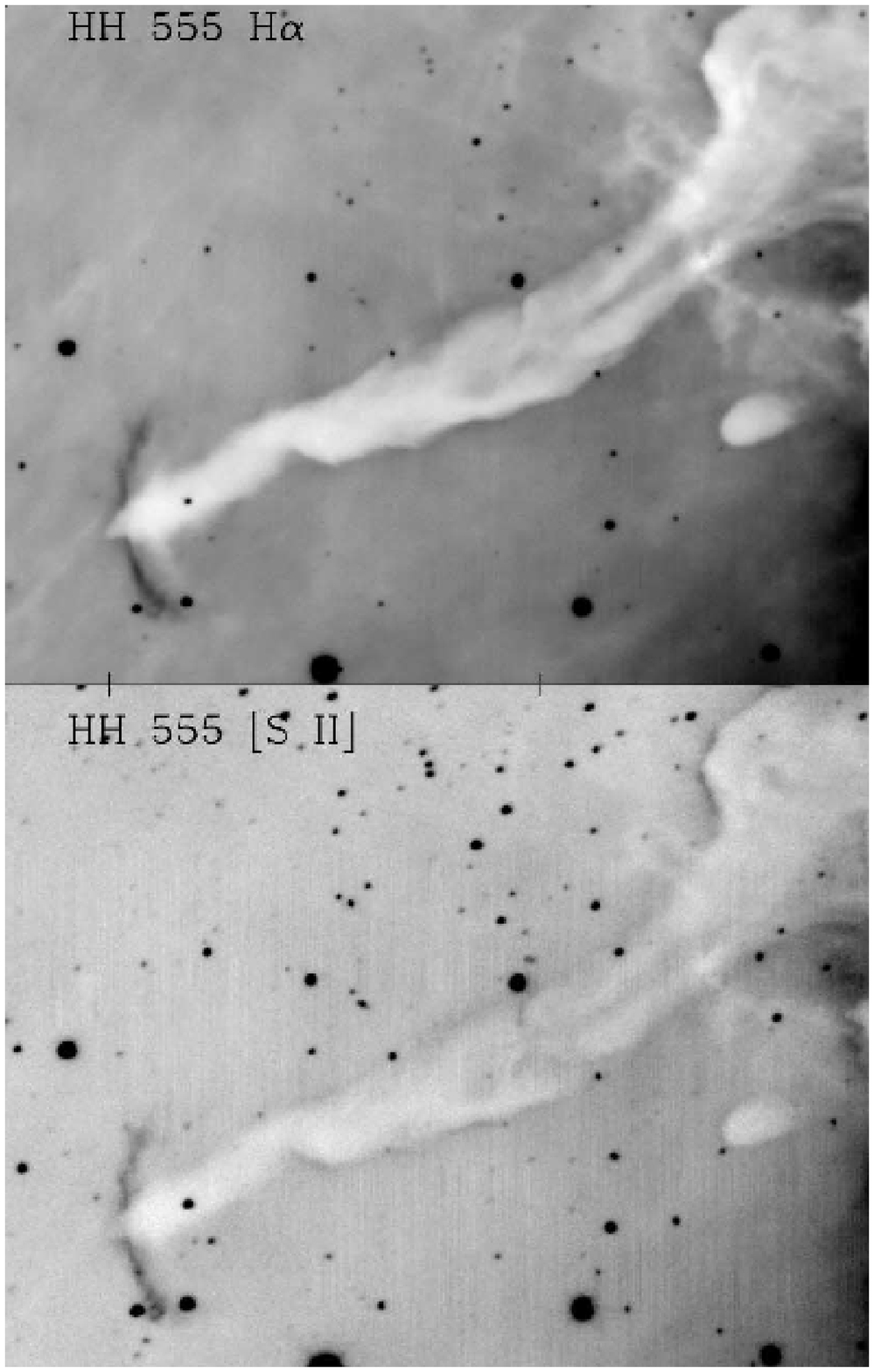}
\caption{A H$\alpha$ and a red [S~II] image of HH~555 taken from Bally \& Reipurth (2003).}
\end{figure}

\begin{figure}
\epsscale{.80}
\plotone{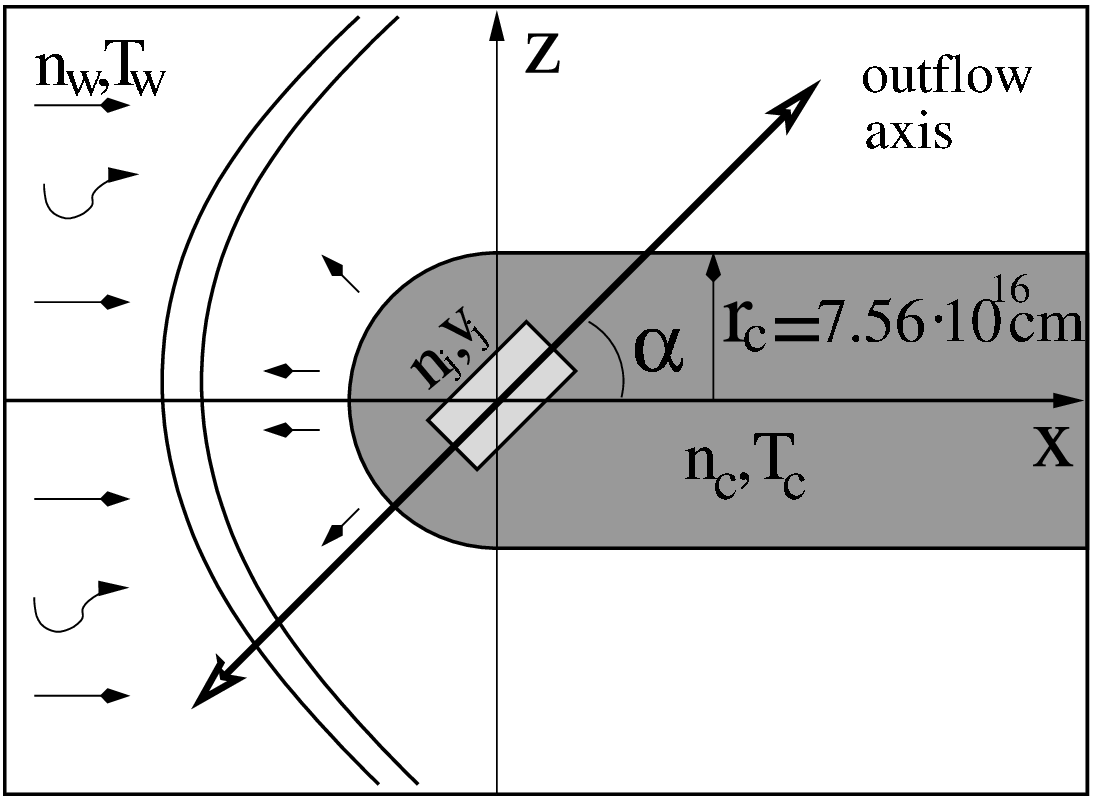}
\caption{A schematic presentation of the numerical setup. The ionizing
photon flux and an ionized, plane-parallel flow are impinging from the left,
along the $x$-axis. The neutral, elephant trunk structure
has the form of a cylinder+spherical cap. From the center of the cap,
a bipolar jet emerges forming an
angle $\alpha$ with the $x$-axis. For a high enough impinging photon
flux, the interaction of the
photoevaporated flow with the impinging plane-parallel wind creates
a two-shock structure around the 
neutral clump.}
\end{figure}

\begin{figure}
\epsscale{0.90}
\plotone{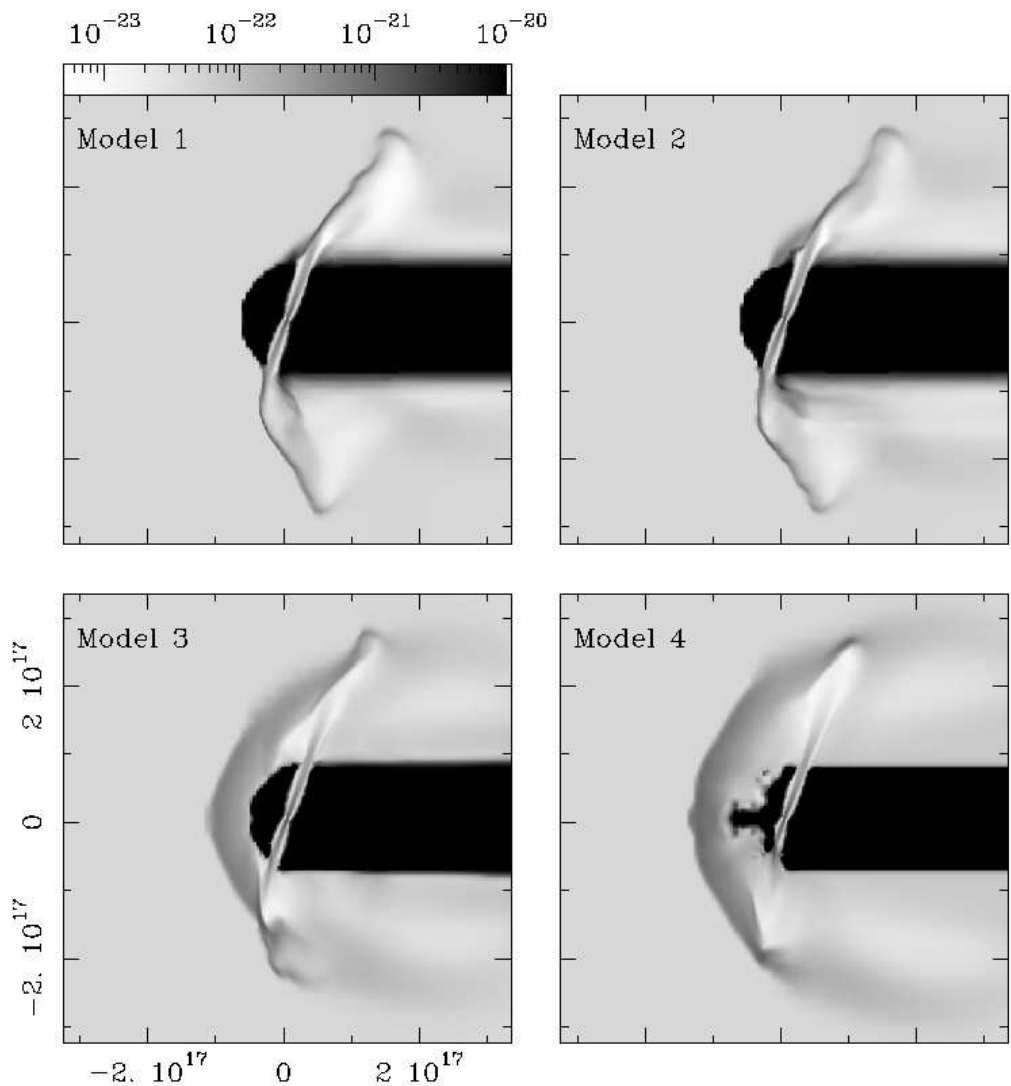}
\caption{Density on the midplane of the simulation obtained from
simulations with different impinging hydrogen-ionizing photon fluxes.
For each model we show a time slice corresponding to a 3000~yr
simulation time. The extent of the computational domain is 
$(6.5,3.25,6.5)\times 10^{17}$~cm along the $(x,y,z)$-axes (respectively)
and the maximum resolution of the adaptive grid is of $1.27\times 10^{15}$~cm.
The photon flux and the expanding H~II region
impinge from the left, parallel to the symmetry axis of the neutral
elephant trunk. The jet is ejected at an angle of  70$^\circ$ 
with respect to the $x$-axis. The photon flux has values
$F_*=(4.2,42,420,4200)\times 10^{7}$~cm$^{-2}$s$^{-1}$ for
models~1 through 4 (respectively). The density stratifications
are depicted with the logarithmic
greyscale shown by the top bar, in units of g~cm$^{-3}$.
The axes are labeled in cm.}
\end{figure}

\begin{figure}
\epsscale{0.90}
\plotone{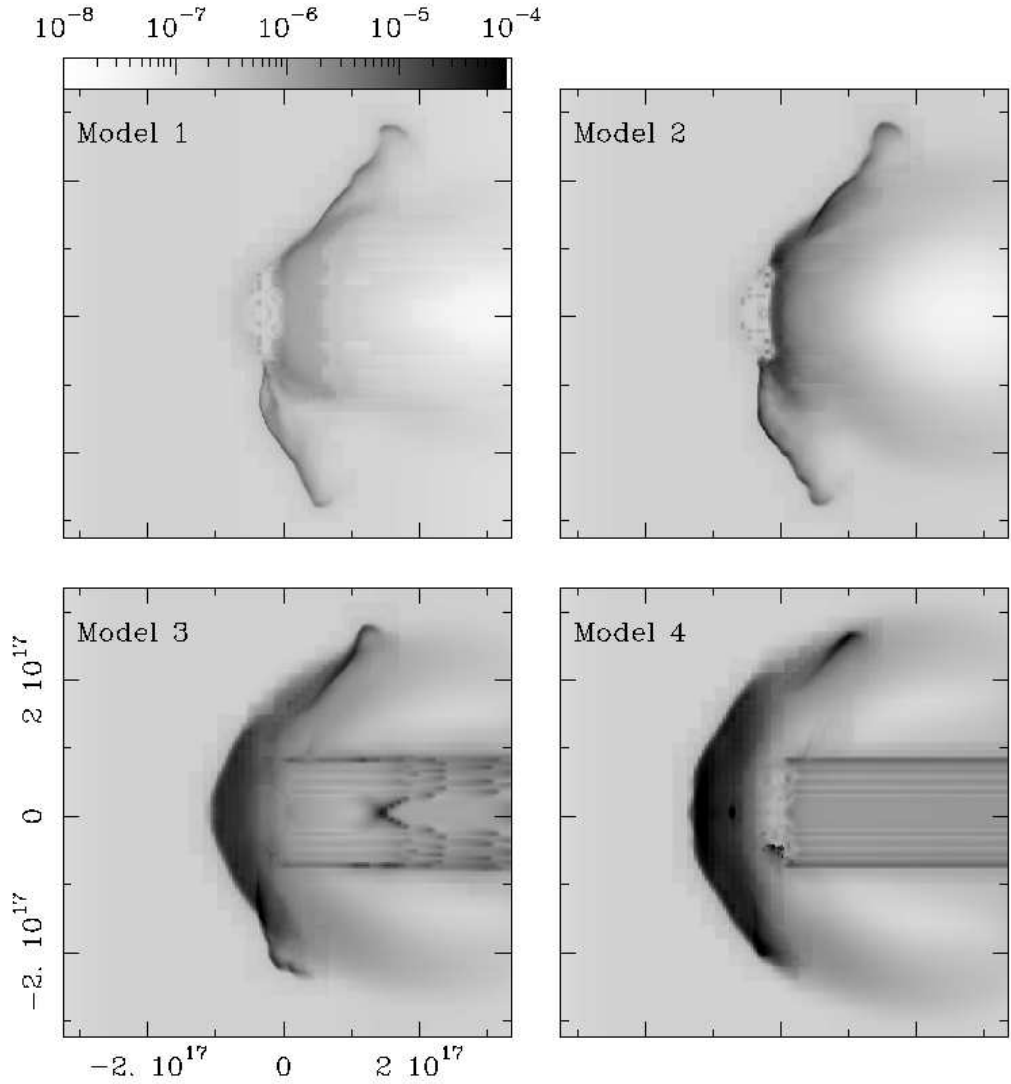}
\caption{H$\alpha$ emission maps obtained
by integrating the H$\alpha$ emission coefficient along the $y$-axis.
The extinction within the elephant trunk
was not considered. The emission is depicted with the logarithmic
greyscale shown by the top bar, in units of
erg~s$^{-1}$~cm$^{-2}$~sterad$^{-1}$. For the details of
the models, see Figure~2 and \S 3.
The axes are labeled in cm.}
\end{figure}

\end{document}